\documentclass[12pt]{article}
\usepackage{epsfig, axodraw}
\usepackage [small] {caption}
\newcommand{\ra}{\rightarrow}
\newcommand{\bq}{\begin{eqnarray}}
\newcommand{\eq}{\end{eqnarray}}

\newcommand{\ov}{\overline}

\begin{document}

\begin{center}{\bf The pion diffractive dissociation into two jets }
\end{center}
\vspace{1cm}

\begin{center} {V.L. Chernyak} \end{center}

\begin{center}{Budker Institute for Nuclear Physics, \\
630090 Novosibirsk, Russia \\
E-mail: chernyak@inp.nsk.su}\end{center}
\vspace{1cm}
\begin{center}
Talk given at the International Workshop: \\ "Exclusive Processes
at High Momentum Transfer", \\ Jefferson Lab, Newport News, VA, USA,
May 15-18, 2002.
\end{center}
\vspace{1cm}
\begin{center}{\bf Abstract}\end{center}

The method used and results obtained are described for calculation of the
cross section for the pion diffractive dissociation into two jets. The main
new qualitative result is that the distribution of longitudinal momenta for
jets is not simply proportional to the profile of the pion wave function
$\phi_{\pi}(x)$,
but depends on it in a complicated way. In particular, it is shown that under
 conditions of the E791 experiment, the momentum distribution of jets is
similar in its shape for the asymptotic and CZ wave functions. It is concluded
therefore that, unfortunately, the process considered is really weakly
sensitive to the profile of the pion wave function, and the accuracy of data
is insufficient to distinguish clearly between different models of
$\phi_{\pi}(x)$.

Comparison with the results of other papers on this subject is given.

\newpage

{\bf 1.} \hspace{1cm} The E791 experiment at Fermilab [1]
has recently measured  the cross section of the hard diffractive
dissociation of the pion into two jets. In particular, the distribution
of the total pion longitudinal momentum into fractions $y_1$ and
$y_2,\, (y_1+y_2)=1,\,$ between jets has been measured. The main purpose
was to obtain in this way the information about the leading twist pion wave
function $\phi_{\pi}(x_1,x_2) $, which describes the distribition of
quarks inside the pion in the longitudinal momentum fractions
$x_1$ and $x_2=1-x_1$.

The hope was based on theoretical calculations of this cross section, see
[2]-[5]. It has been obtained in these papers that the
cross section is simply proportional to the pion wave function
squared: $d\sigma/dy_1\sim |\phi_{\pi}(y_1)|^2$. In such a case, it
would be sufficient to measure only the gross features of $d\sigma/dy$
to reveal the main characteristic properties of $\phi_{\pi}(x)$, and
to discriminate between various available models of $\phi_{\pi}(x).$

Our main qualitative result [6]-[7] is that this is not the case.
The real situation is much more complicated, with
$d\sigma/dy$ depending on $\phi_{\pi}(x)$ in a highly nontrivial way.
We describe below in a short form our approach and the results obtained
for this cross section.

{\bf 2.} \hspace{1cm} The kinematics of the process is shown in fig.1.
 We take the nucleon as a target, and the initial and final nucleons
are substituted by two soft gluons with momenta $q_1=(u+\xi){\ov P}$
and $q_2=(u-\xi){\ov P},\, {\ov P}=(P+P^{\prime})/2,\, \Delta=(q_1-q_2).$
\footnote{
The small skewedness, $\xi \ll 1$, is always implied. It is typically:
$\xi \sim 10^{-2}$, in the Fermilab experiment.}
The lower blob in fig.1 represents the generalized
gluon distribution of the nucleon, $G_{\xi}(u)=G_{\xi}(-u)$ [8]-[10].

The final quarks are on shell, carry the fractions $y_1$ and $y_2$ of the
initial pion momentum, and their transverse momenta are:
 $({\bf k}_{\perp}+({\bf q}_{\perp}/2))$ and $(-{\bf k}_{\perp}+
({\bf q}_{\perp}/2)),\,  q_{\perp}\ll k_{\perp}$, where
$ q_{\perp}$ is the small final transverse momentum of the target,
while $ k_{\perp}$ is large. The invariant mass of these two quark jets is:
$M^2=k_{\perp}^2/y_1y_2$, and $\xi=M^2/\nu,\,\,\nu=(s-u).$

The upper blob M in fig.1 represents the hard kernel of the amplitude.
According to the well developed approach to  description of
hard exclusive processes in QCD [11]-[14] (see [15] for a review), all
hard gluon and quark lines in all diagrams
have to be written down explicitly and substituted
by their perturbative propagators. In other words, the hard momentum flow
have to be made completely explicit and these hard lines of diagrams
constitute the hard kernel M. They should not be hidden (if it is
possible at all)
as (a derivatives of) "the tails" of the unintegrated pion wave function
$\Psi_{\pi}(x,\,l_{\perp})$, or of the "unintegrated gluon distribution".
This is, first of all, what differs our approach from previous
calculations of this process in [4]-[5].

For calculation of M in the leading twist approximation and in the lowest
order in $\alpha_s$, the massless pion can be substituted
by two massless on shell quarks with the collinear momenta
$x_{1}p_{\pi}$ and $x_{2}p_{\pi}$ and with zero transverse momenta, as
account of primordial virtualities and transverse momenta results only in
higher twist corrections to M. The leading twist pion wave function
$\phi_{\pi}(x,\mu_o)$ describes the distribution of these quarks in
momentum fractions $x_1$ and $x_2$.
\footnote{
As usual, on account of leading logs from loops the soft pion wave function
$\phi_{\pi}(x,\mu_o)$ evolves to $\phi_{\pi}(x,\,\mu)$,
where $\mu$ is the characteristic scale
of the process. In other words, in the leading twist component of the pion
wave function the two pion quarks prepare themselves for a
hard collision by exchanging gluons and increasing their virtualities
and transverse momenta from the initial scale $\mu_o\sim \Lambda_{QCD}$ up
to the characteristic scale of the process
$\mu \leq k_{\perp}$, so that this
quark pair enters nucleon (nuclei) having already the small transverse size
$r_{\perp}\sim 1/{k_{\perp}}$ (while the smallness of the longitudinal size
is ensured by the Lorenz contraction).
}

So, the hard kernel M is proportional to the scattering
amplitude of two initial collinear and on shell quarks of the pion on the
on shell gluon:
\bq
M\sim \left \{ d(x_{1}p_{\pi})+{\bar u}(x_{2}p_{\pi})+g(q_{1})\ra
d(p_1)+{\bar u}(p_2)+g(q_2)\right \}\,\,\,.
\nonumber
\eq
In lowest order in $\alpha_s(k_{\perp})\,\,\,$  M consists of 31
connected Born diagrams, each one is $\sim O(\alpha_{s}^{2}(k_{\perp}))$.

The general structure of the amplitude is, therefore (symbolically):
\bq
T\sim \langle P^{\prime}|A^{\perp}\cdot A^{\perp}|P \rangle \otimes({\bar
\psi}_1 M \psi{_2}) \otimes \langle 0|{\bar u}\cdot d|\pi^{-} \rangle,
\nonumber
\eq
where the first matrix element introduces the skewed gluon distribution
of the nucleon $G_{\xi}(u)$,
${\bar \psi}_1$ and $\psi{_2}$ are the free spinors of final
quarks, "M" is the hard kernel, i.e. the product of all vertices and hard
propagators, the last matrix element introduces the pion wave function
$\phi_{\pi}(x)$, and $\otimes$ means the appropriate convolution.

\begin{figure}
\centering
\begin{picture}(300,180)(0,0)
    \SetColor{Black}
    \ArrowLine(10,125)(55,125)
    \Text(30,140)[]{\large $p_\pi$}
    \COval(75,125)(30,20)(0){Black}{White}
    \Text(75,125)[]{\large $\phi_\pi(x)$}
    \ArrowLine(80,95)(150,95)
    \Text(110,85)[]{\large $x_2 p_\pi$}
    \ArrowLine(80,155)(150,155)
    \Text(110,165)[]{\large $x_1 p_\pi$}
    \ArrowLine(195,95)(230,80)
    \Text(245,80)[]{\large $p_2$}
    \ArrowLine(195,150)(230,165)
    \Text(245,165)[]{\large $p_1$}
    \DashArrowLine(150,40)(150,90){5}
    \Text(140,65)[]{\large $q_1$}
    \DashArrowLine(190,90)(190,40){5}
    \Text(200,65)[]{\large $q_2$}
    \ArrowLine(110,10)(140,25)
    \Text(100,15)[]{\large $p$}
    \ArrowLine(200,25)(230,10)
    \Text(240,15)[]{\large $p'$}
    \COval(170,125)(40,40)(0){Black}{White}
    \Text(170,125)[]{\large $M$}
    \COval(170,25)(20,30)(0){Black}{White}
    \Text(170,25)[]{\large $G_\xi(u)$}
\end{picture}
\caption{Kinematics and notations}
\label{Fig.1}
\end{figure}
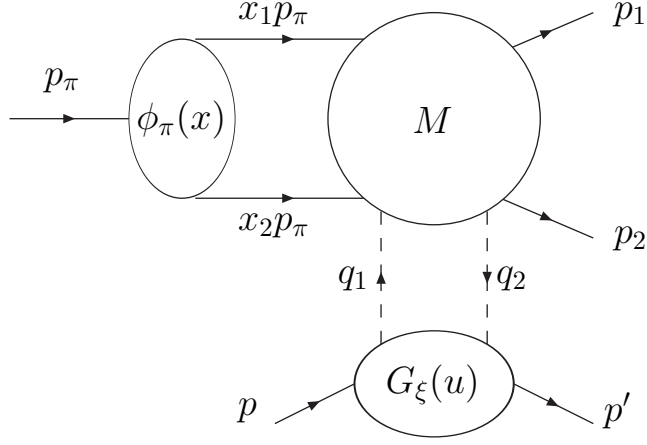

{\bf 3.} \hspace {1cm} Proceeding in the above described way and summing up
contributions of all Born diagrams, one obtains for the cross section
\footnote{
Only the imaginary part of the amplitude, $Im\, T_g$ (from the gluon
distribution in the target), is shown explicitly in Eqs.(1)-(6), as
it is expected to be the main one at high energies. For obtaining $Im\, T$,
the terms $i\epsilon$ were introduced into all denominators through:
$s\ra s+i\epsilon$, i.e.
$\xi\ra \xi-i\epsilon$ [6]. The contribution of the quark distribution to
$Im\, T$ is given explicitly in [7]. Let us note also that all integrals
over $"x_1"$ entering Eqs.(2)-(6) are convergent at the ends points
$x_1\ra 0$ or $x_1\ra 1$.
}
:
\bq
d\sigma_N=\frac{1}{8\,(2\pi)^5}\frac{1}{s^2}\,|T|^2\,\frac{dy_1}{y_1 y_2}
d^2 k_{\perp}d^2 q_{\perp};\,\,\,
Im\,T_g=\frac{2\pi\,s\,\omega_o}{k_{\perp}^2}\,G_{\xi}(\xi)\,\Omega,
\eq
\bq
\Omega=\int_0^1 dx_1\,\phi_{\pi}(x_1)\left (\Sigma_1+\Sigma_2+\Sigma_3+
\Sigma_4\right )\,,
\eq
\bq
\Sigma_1=\left [\,\frac{4}{x_1 x_2\,|x_1-y_1|}\,\frac{G_{\xi}({\bar u})}
{G_{\xi}(\xi)}\,\Theta(|x_1-y_1|> \delta)\,\right ]+(y_1 \leftrightarrow y_2),
\eq
\bq
\Sigma_2=\frac{1}{x_1^2\,x_2^2\,y_1\,y_2}\Biggl \{\,- (x_1 x_2+y_1 y_2)+
\Biggr. \nonumber
\eq
\bq
\Biggl. +\left [ |x_1-y_1|(x_1-y_2)^2\,\frac{G_{\xi}({\bar u})}
{G_{\xi}(\xi)}\,\Theta(|x_1-y_1|> \delta) +
(y_1\leftrightarrow y_2) \right ] \Biggr \}\,,
\eq
\bq
\hspace{-5mm}{
\Sigma_3=\frac{x_1 x_2+y_1 y_2}{9 x_1^2 x_2^2 y_1 y_2}
\Biggl \{-1+\left [|x_1-y_1|
\frac{G_{\xi}({\bar u})}{G_{\xi}(\xi)}\Theta(|x_1-y_1|>\delta)
+(y_1 \leftrightarrow y_2)\right]\Biggr \}
}
\eq
\bq
\Sigma_4=\,\frac{16}{9}\,\frac{1}{x_1 x_2 y_1 y_2}\,\xi\,\frac{
{dG_{\xi}(u)/du}|_{u=\xi}}{G_{\xi}(\xi)}\,;\,\,\,\,
\quad \delta=\frac{k_{\perp}^2}{s},
\eq
\bq
\omega_o=\frac{\delta_{ij}(4\pi\alpha_s)^2 f_{\pi}}{96}({\ov \psi}_1
\Delta_{\mu}\gamma_{\mu}\gamma_5\psi{_2})\,\frac{(y_1 y_2)^2}{k_{\perp}^4};
\,\,\,\,{\bar u}=\xi\left(\frac{x_1 y_2+x_2 y_1}{x_1-y_1}\right ).
\eq

It is seen from the above equations that $d\sigma/dy$ is not $\sim |\phi_
{\pi}(y)|^2$, but depends on the profile of $\phi_{\pi}(x)$ in a very
complicated way.

{\bf 4.} \hspace{1cm} In this section we present some numerical estimates
of the cross section, based on the above expressions (1)-(6). Our  main
purpose was to trace the distribution of jets in longitudinal momentum
fractions $y_1,y_2$ depending on the profile of the pion wave function
$\phi_{\pi}(x).$

{\bf a}) For the skewed gluon distribution $G_{\xi}(u,t,\mu)$ of
the nucleon at $t\simeq - q_{\perp}^2\simeq 0$
we used the simple form (as we need it at $|u|\geq \xi$ only, and because
$G_{\xi}(u)\ra G_o(u)$ at $|u|\gg \xi$):
\bq
G_{\xi}(u,t=0,\,\mu \simeq k_{\perp}\simeq 2\,GeV)|_
{u\geq \xi}\simeq u^{-0.3}(1-u)^5.
\eq
This form agrees numerically reasonably well with
the ordinary, $G_o(u,\mu\simeq 2\,GeV)$, and skewed, $G_{\xi}(u,t=0,
\mu\simeq 2\,GeV)$, gluon distributions of the nucleon calculated
in [16] and [17] respectively (in the typical region
of the E791 experiment: $|u|\geq \xi \sim 10^{-2}$).

{\bf c}) As for the pion leading twist wave function, $\phi_{\pi}(x,\,\mu)$,
we compare two model forms: the asymptotic form, $\phi_{\pi}^{asy}(x,\,
\mu)=6x_1x_2$, and the CZ-model [18]. The latter has the form:
$\phi_{\pi}^{CZ}(x,\,\mu_o\simeq 0.5\,GeV)=30\,x_1x_2(x_1-x_2)^2$, at the
low normalization point. Being evolved to the characteristic scale of
this process, $\mu \simeq k_{\perp}\simeq 2$\,GeV, it looks as:
$\phi_{\pi}^{CZ}(x,\,\mu\simeq 2\,GeV)=15\,x_1x_2\Bigl [(x_1-x_2)^2+0.2
\Bigr ],\,$ see Fig.2.

The results of these numerical calculations are then compared with the
E791-data, see Fig.3.

\begin{figure}
\centering
\includegraphics[width=0.5\textwidth]{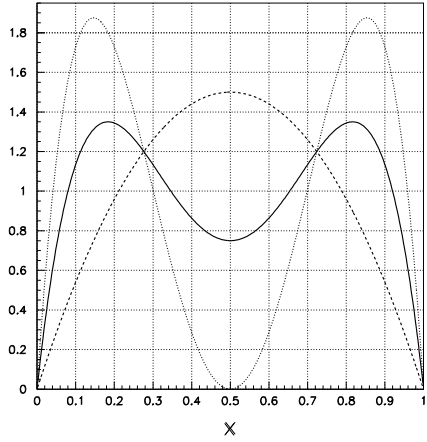}
\vspace{-7mm}
\caption{Profiles of the pion wave functions:
a) $\phi_{\pi}^{CZ}(x,\mu \simeq 0.5\,GeV)=30\,x_1x_2(x_1-x_2)^2$ -
dotted line;\,
b) $\phi_{\pi}^{CZ} (x, \mu\simeq 2\,GeV)=15\,x_1x_2[0.2+(x_1-x_2)^2]$
- solid line;\, c) $\phi_
{\pi}^{asy}(x)=6\,x_1x_2$ - dashed line.}
\label{Fig.2}
\end{figure}~

\begin{figure}
\centering
\includegraphics[width=0.5\textwidth]{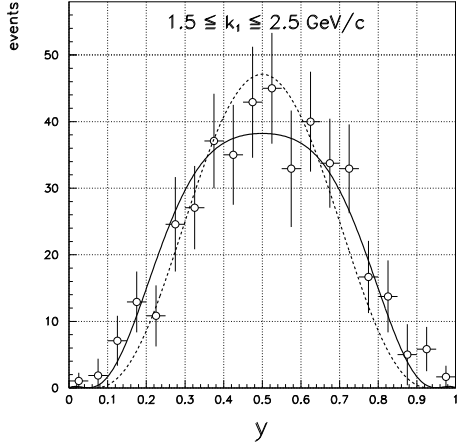}
\vspace{-7mm}
\caption{The y-distribution of jets calculated for $k_{\perp}=2\,GeV,\,
E_{\pi}=500\,GeV$ and with the pion wave functions: $\phi_{\pi}^{CZ}(x,\,
\mu\simeq 2\,GeV)$ - solid line,\, $\phi_{\pi}^{asy}(x)$ - dashed line. The
overall normalization is arbitrary, but the relative normalization of
two curves is as calculated. The data points are from the E791
experiment [1].}
\label{Fig.3}
\end{figure}

\newpage

It is seen that while the two model pion wave functions are
quite different, the resulting distributions of jets in longitudinal momenta
are similar and, it seems, the present experimental accuracy is insufficient
to distinguish clearly between them. Moreover, even the ratio of the
differential cross sections is not much different from unity:
$d\sigma^{asy}/d\sigma^{CZ} \simeq 1.2$ at $y_1=0.5$, and the same
ratio is $\simeq 0.7$ at $y_1=0.25$.
\footnote{
On account of the quark contributions to $Im\, T$ the resulting curves for
$d\sigma/dy$ remain very similar to those in Fig.3, see [7]. For the
numerical calculation of $Re\, T$ see [20]. The estimate of $Re\,T$ can
be obtained from the well known formula: $Re\,T(\nu)\simeq (\pi/2)\, d\ln (
\nu^{-1}\,Im\,T(\nu))/d\ln \nu$. Because $\nu^{-1}\,Im\,T(\nu)\sim \nu^{0.3}$
here, $Re\,T\simeq 0.45\,Im\,T$ and $|T|^2\simeq 1.2\,(Im\,T)^2$.
}

{\bf 5.} \hspace{1cm} The purpose of this section is to compare (in a short
form) the above results and those obtained by other authors, see [4], [5] and
[19]-[20].

{\bf a)}\quad In comparison with Eqs.(1)-(6),
the approximations used in [5] correspond to
neglecting the terms $\Sigma_2,\,\Sigma_3$ and $\Sigma_4$, and supposing the
region of small $|x_1-y_1|$ dominates the only remaining contribution:
$\int dx_1\phi_{\pi}(x_1)
\Sigma_1\simeq \phi_{\pi}(y_1)\int dx_1\Sigma_1$. In this approximation,
indeed, $Im\,T_g\sim \phi_{\pi}(y)$ and $d\sigma/dy\sim |\phi_{\pi}(y)|^2$.

Unfortunately, at the real conditions of the E791 experiment this is a poor
approximation, and on account of neglected terms in Eqs.(1)-(6) the form of
$d\sigma/dy$ changes ${\it qualitatively}$ in comparison with
$|\phi_{\pi}(y)|^2$, see Figs.2 and 3.~
\footnote{
The forms of $d\sigma/dy$ for $\phi_{\pi}^{asy}(x)$ and $\phi_{\pi}^{CZ}(x,
\mu\simeq 2\,GeV)$ and with the pion energy ten times larger, $E_{\pi}
=5\,TeV,\,k_{\perp}=2\,GeV$, are shown in Fig.6 in [6].
It is seen from therein that even this does not help much (as the form of the
distribution in $y$ of $d\sigma/dy$ weakly depends on the pion energy), and so
the approximation used in [5] remains poor even at such energies.
}

{\bf b)}\quad The authors of papers in [4] argued that from the whole set of
31 Feynman diagrams only a small especially chosen subset dominates
$Im\,T_g$, while contributions of all other diagrams will be suppressed by
Sudakov form factors. From this subset of diagrams they obtained also:
$Im\,T_g(y)\sim \phi_{\pi}(y)$.

First, I am unable to understand this last result, as the direct calculation
of this subset of diagrams according to the rules described above in sect.2
shows that, here also, resulting  $Im\,T_g(y)$ depends on $\phi_{\pi}(x)$
through a complicated integrated form, like those in Eqs.(4)-(6) above.

As for the Sudakov suppression of all other diagrams, I also disagree. The
line of reasoning in [4] was the following. Consider, for instance, the
diagrams like those in Fig.2 in [6]. Exchanging the hard gluon, the two pion
quarks undergo abrupt change of their direction of motion and will tend to
radiate gluons in their previous direction of motion. Because the events
with such radiation are excluded from the data set, what remains will be
suppressed by the Sudakov form factors.

Applying this line of reasoning to nearly any hard exclusive amplitude, one
will conclude that all such amplitudes will be suppressed, {\it as a whole},
by Sudakov effects. For instance, let us consider the large angle Compton
amplitude. In the c.m.s. and at high energy, the three proton quarks change
abruptly their direction of motion, deviating by large angle $\theta$. So,
following the line of reasoning in [4], the whole large angle Compton
amplitude will be suppressed by the Sudakov form factors.

Really, this is not the case. The reason is that the radiations of three
collinear proton quarks cancel each other due to colour neutrality of the
proton.
\footnote{
This cancelation becomes ineffective only at the ends points of the
quark phase space where, for
instance, one quark carries nearly all the proton momentum, while other
ones are wee. The contributions from these regions to the total
amplitude are, indeed, suppressed by the Sudakov effects. But in any
case, the contributions from these end point regions give only power
corrections to the total amplitude, even ignoring the additional Sudakov
suppression.}

{\bf c)}\quad Unlike [4] and [5], the results from [19]-[20] are very
similar to those from [6]-[7]. In particular, the analytic expressions
for $(T_g+T_q)$ in terms of corresponding integrals over $\phi_{\pi}(x)$
and the gluon (quark) distributions are the same, the only difference is
that the signs of $i\epsilon$ in denominators of a few terms are opposite.
For $T_g$ for instance (and similarly for $T_q$), this difference can be
represented as:
\bq
\hspace{-2mm}{
T_g=\delta T_g (\nu+i\epsilon, M^2)+
\int_0^1 dx_1 \phi_{\pi}(x_1)[ f_1 (x,y) g^{+}(\nu, M^2)+
f_2 (x,y) g^{\pm}(\nu, M^2) ],\nonumber
}
\eq
\bq
g^{\pm}(\nu, M^2)=\int_{-1}^1 du\frac{G_{\xi}(u)}{(u+
\frac{M^2}{\nu\pm i\epsilon})(u-\frac{M^2}{\nu\pm i\epsilon})}\,\,\,\,,
\eq
where $\delta T_g$ is the same in [6] and [19]-[20],
and $f_1(x,y)$ and $f_2(x,y)$ are definite simple
functions. The upper sign in Eq.(9) is from [6] (where as pointed out
above, the terms $i\epsilon$ were introduced into all denominators through:
$\nu\ra\nu+i\epsilon$), while the lower sign is from [19]-[20].

The result for $Im\,T_g$ was obtained in [19]-[20] by direct calculation of
one-loop Feynman diagrams, and so it can be checked only by a similar
independent calculation. But there at least one simple argument
following from
general analyticity properties, in favour of the approach used in [6].

Let us first consider our amplitude $T_g(\nu, M^2)$ in the Euclidean region
$M^2 < 0$. There is no discontinuity in $M^2$ in this region, and the only
discontinuity is due to the s-cut. So, $i\epsilon$ terms enter unambiguously
into {\it all} denominators as: $\nu=(s-u)\ra \nu+i\epsilon$,
and the terms $f_1$ and $f_2$ in Eq.(9) will be multiplied by
{\it the same analytic function} $g(\nu, M^2)$:
\bq
[\,f_1(x,y)+f_2(x,y)\,]\,g(\nu+i\epsilon, M^2)\,\,.
\eq

Let us continue now this expression into the Minkowski region $M^2 > 0$. How
it may be that after the analytic continuation the functions $f_1(x,y)$ and
$f_2(x,y)$ will be multiplied by two different functions, as in [19]-[20]
(see Eq.(9)):
\bq
[\,f_1(x,y)\,g(\nu+i\epsilon, M^2)+f_2(x,y)\,g^{\star}(\nu+i\epsilon,
M^2)\,]\quad ?
\eq
\vspace{5mm}

{\bf In conclusion}, I would like to emphasize that in spite of some
"internal" differences between the results from [6]-[7] and [19]-[20],
the final curves for $d\sigma/dy$ are practically the same, and Fig.3
gives a good representation of both answers. So,
in any case, the qualitative
conclusions agree: under the conditions of the E791 experiment, the process
considered appeared to be weakly sensitive to the profile of the pion wave
function $\phi_{\pi}(x)$ and, unfortunately, these data can not
discriminate between, say, $\phi_{\pi}^{asy}(x)$ and $\phi_{\pi}^{CZ}(x)$.

\vspace{4mm}
\begin{center}\bf Acknowledgements \end{center}

I am grateful to the Theory Group of JLab and Organizing Committee and,
in particular, to A.V. Radyushkin for a kind hospitality and support.

\end{document}